\documentclass[12pt]{article}
\usepackage{mathbbold}
\usepackage{epsfig}

\def\beq{\begin{equation}}
\def\eeq#1{\label{#1}\end{equation}}
\def\eeqn{\end{equation}}

\def\beqa{\begin{eqnarray}}
\def\eeqa#1{\label{#1}\end{eqnarray}}
\def\eeqan{\end{eqnarray}}

\let\bar=\overbar

\def\Dslash{\not{\hbox{\kern-4pt $D$}}}
\def\dslash{\not{\hbox{\kern-2pt $\del$}}}

\def\msb{{\bar{\ssstyle M \kern -1pt S}}}

\def\s#1{\widetilde{#1}}

\usepackage{fancyhdr,graphicx}
\def\Title#1{\begin{center} {\Large {\bf #1} } \end{center}}

\begin{document}

\Title{The Deconfinement Phase Transition in the Interior of Neutron Stars}

\bigskip\bigskip


\begin{raggedright}

{\it Xia Zhou$^{1,2}$, Xiao-Ping Zheng$^1$\\
$^1$ Institute of Astrophysics\\
 Huazhong Normal University\\
 Wuhan 430079\\
 P.R.China.\\

 $^2$ Urumqi Observatory, NAOC, CAS,40-5 South BeiJiang Road, Urumqi, 830011, P.R.China.\\
{\tt Email: zhouxia@nao.ac.cn}}
\bigskip\bigskip
\end{raggedright}

\begin{abstract}
The deconfinement phase transition which happens in the interior of neutron stars are investigated. Coupled with the spin evolution of the stars, the effect of entropy production and deconfinement heat generation during the deconfinement phase transition in the mixed phase of the neutron stars are discussed. The entropy production of deconfinement phase transition can be act as a signature of phase transition, but less important and does not significantly change the thermal evolution of neutron stars. The deconfinement heat can change the thermal evolution of neutron star distinctly.

\end{abstract}

\section{Introduction}
The interior of neutron stars are considered as systems where high-density phase of strong interacting matter do exist. The chemical composition of neutron stars at densities beyond nuclear saturation densities ranging from purely nucleonic matter through hyperon or meson condensation to deconfined quark matter \cite{ref1,ref2}. Comparing with theoretical models with the observational data, we have opportunity to constrain or understanding fundamental elements of particle and nuclear physics. Many works believed that the appearance of quark matter is implied in the structure and evolution of neutron stars \cite{ref3,ref4,ref5,ref6,ref7,ref8,ref9}. The consequences of different phase scenarios and spin evolution for the thermal evolution of neutron stars have been reviewed in comparison with existing soft x-ray observation of thermal radiation from neutron star surfaces. They both depend sensitively on and to some extent determine the chemical composition of the stars \cite{ref10,ref11,ref12,ref13}.

As neutron stars spin down and contract, their structure and composition change with the increasing density, and the quark matter may appear. We will discuss how the increasing density affect the spin and thermal evolution of neutron stars which contain quark matter in the core. As mentioned in the work \cite{ref17}, the increasing density and changing chemical composition further imply additional entropy production in bulk and the release of the latent heat as particles cross any phase boundaries present. They also though that this kin of additional entropy production during deconfinement phase transition should affect the spin and thermal evolution of neutron stars.

In our former work, we discussed that the occurrence of first-order deconfinment phase transition is accompanied by the release of energy which originate from the binding energy. Therefore, we introduced the deconfinement phase transition which happened in the mixed phase of neutron star in this paper. The energy released during the deconfinement phase transition can be calculated with different equation significantly \cite{ref10,ref11,ref12,ref15,ref16}.

In Sect. $2$ we will review the deconfinement phase transition which happens in the core of neutron stars, compare entropy production and deconfinement heat. Its application to the spin and thermal evolution of neutron stars also will be discussed. The conclusion will be given out in Sect. $3$.

\section{Deconfinement phase transition}
As the star spins dowm, the centrifugal force decreases continuously, increasing its internal density. At this current, the nuclear matter continuously converts into quark matter in an exothermic reaction, i.e. $n \rightarrow u + 2d$, $p \rightarrow 2u + d$, $s$ quarks immediately appear after weak decay. Quarks are accumulating in the interior of the neutron star with decreasing rotation frequency.

In the work of M. Stejner et al. \cite{ref17}, the authors introduced a possible connection between the spin down and thermal evolution of the neutron stars with a deconfinement phase transition which originates from the spin-down-powered entropy production might affect the thermal evolution of the stars. The signature of the appearance of a pure quark matter core is related to the change in radius, chemical composition, structure and under certain circumstances the additional entropy production both in bulk and in the form of latent heat. For a star of constant density and temparture with a baryon number of $10^{56}$. A reasonable approximation of the total additional entropy production can be written as \cite{ref17}:
\begin{equation}
H_{L}\sim10^{41} T^{2}_{9}(\frac{\rho}{fm^{-3}})^{-1/3}(\frac{B}{10^{14}G})^2(\frac{\Omega}{6000rad
s^{-1}})^4 erg\\s^{-1}
\end{equation}
where $\rho$ is the baryon number density of the stars.

For neutron stars in which mixed phase exist, this results in the
first-order deconfinement phase transition with varying pressure.
The charge of enthalpy over the transition region immediately leads
to the release of energy. The released energy during such phase
transition can heating the neutron stars at old ages. The work
of~\cite{ref14} give out the calculation of energy release per
baryon during such phase transitions. The thermal evolution of
neutron star with deconfinement heating also have been discussed
in~\cite{ref12}. Since the mean energy release per converted baryon
has been obtained, heating luminosity of deconfinement heating can
be written as~\cite{ref14}
\begin{equation}
H_{D}=N\delta\dot{q}\sim-N\bar{q}\frac{2\Omega\dot{\Omega}}{\Omega^2_\lambda}\sim10^{45}(\frac{\bar{q}}{0.1MeV})(\frac{B}{10^{14}G})^2(\frac{\Omega}{6000rad
s^{-1}})^4 erg s^{-1}
\end{equation}
where $\bar{q}$ is the mean energy release per converted baryon.

With the help of different observations, a fraction of observable
pulsar's thermal radiations were detected in soft X-ray and UV
bands. These pulsars are middle aged ($10^3~10^6yr$) with temperature
about $0.3-1MK$ \cite{ref18}. When the compact star is older than
$10^6yr$, the photo cooling dominated neutrino emission and the stars
cool down rapidly. The temperature of such stars is too low to be
measured through the thermal radiation based on traditional thermal
evolution theoties. PSR J0437-4715 has been detected in UV/FUV with
HST~\cite{ref19}. The shape of the inferred spectrum suggests
thermal emission from the whole neutron star surface of a
surprisingly high temperature of about $10^5K$. A powerful energy
source should be operating in a Gyr-old neutron star to keep its
surface at such high temperature. The heating effects are expect to
be significant for old compact stars. Here, we will compare the two
different energy generation mechanism with the observational data of
PSR J0437-4715.

For PSR J0437-4715 with $P=5.76ms,\dot{P}=5.73\times10^{-20}$,
heating luminosity of deconfinement heating $H_D~10^{31} erg
s^{-1}$. This estimate under no consideration of effects of
space-time curvature is slight higher than the inferred thermal
X-ray luminosity $L_x~10^{29}erg S^{-1}$~\cite{ref10}. The entropy
production can be estimated to be $H_L$ $10^{27}\times T^2_{9}$ $ erg
s^{-1}$ which is determined by the temperature of stars.As mentioned
in the work~\cite{ref17},the signature of deconfinement found here
is below the present observational sensitivity and not of sufficient
strength to set apart the thermal evolution curves with temperature
versus time for neutron stars.

\section{Discussion and conclusions}

  The deconfinement phase transition during the spin-down of neutron
stars have been discussed in this paper.We estimated the energy
generation rate of the two different mechanisms during deconfinement
phase transition.
  The entropy release during deconfinement phase transition which
  analyzed in the work~\cite{ref17} have found do not dominate the general
  thermal evolution of neutron star but they do complement the standard picture.
  The signature of deconfinement found
in this work is below the present observational sensitivity and not of sufficient
strength to set apart the cooling curves with temperature versus time for hybrid
stars.

The energy generated from first-order phase transitions which can act as a heating
mechanism during the thermal evolution of neutron stars. It can affect the thermal
evolution of compact star significantly at older age.

But the equation of state of superdense matter is still a mystery and the stars
distance and ages are uncertain. Therefore, constraint on the chemical composition
of compact stars purely through thermal radiation will be limited. Other observed
phenomena, such as critical rotation and limited mass-radius relationship
\cite{ref20,ref21,ref22,ref23,ref24,ref25} should be used for this goal. The combination of
different evolution processes, such as spin and thermal evolution of the stars which is
mentioned in \cite{ref17}, is a useful way to understand superdense matter in the compact
star.

\bigskip
This work is supported by the NSFC project NO. 10773004 and the 973Program 2009CB824800.

\def\Discussion{
\setlength{\parskip}{0.3cm}\setlength{\parindent}{0.0cm}
     \bigskip\bigskip      {\Large {\bf Discussion}} \bigskip}
\def\speaker#1{{\bf #1:}\ }
\def\endDiscussion{}

\Discussion

\speaker{D. Boss (University of Emperor)} The result is not trustable, since we
are transmiting troops using frequency of 0.31 Hz.

\speaker{Med} Professor Boss has discussed the possibility of signal mixture,
however we have EM shielding.

\endDiscussion

\end{document}